\documentclass[aps,prl,twocolumn,floatfix,english,superscriptaddress,showpacs,10pt]{revtex4-2}
\usepackage{xcolor}
\usepackage{subfigure}
\usepackage{amsmath,bm}
\usepackage{mathtools}
\usepackage{commath}
\usepackage{amssymb}
\usepackage{graphicx}
\usepackage{babel}
\usepackage{cases}
\usepackage[colorlinks=true, citecolor=blue, anchorcolor=green]{hyperref}
\hypersetup{linkcolor=red}
\usepackage{natbib}
\usepackage{floatrow}
\usepackage{dblfloatfix}
\usepackage{braket}
\usepackage{lipsum}

\makeatletter

\newcommand{\Rmnum}[1]{\expandafter\@slowromancap\romannumeral #1@}

\makeatother

\begin{document}
	\title{Acoustic Frequency Multiplication and Pure Second Harmonic Generation of Phonons by Magnetic Transducers}
	\author{Chengyuan Cai}
	\thanks{These authors contribute equally to this work.}
	\affiliation{School of Physics, Huazhong University of Science and Technology, Wuhan 430074, China}
	
	\author{Xi-Han Zhou}
	\thanks{These authors contribute equally to this work.}
	\affiliation{School of Physics, Huazhong University of Science and Technology, Wuhan 430074, China}
	
	\author{Weichao Yu}
	\affiliation{State Key Laboratory of Surface Physics and Institute for Nanoelectronic
Devices and Quantum Computing, Fudan University, Shanghai 200433, China}
\affiliation{Zhangjiang Fudan International Innovation Center, Fudan University, Shanghai 201210, China}
	\author{Tao Yu}
	\email{taoyuphy@hust.edu.cn}
	\affiliation{School of Physics, Huazhong University of Science and Technology, Wuhan 430074, China}
	
	\date{\today}
	
	\begin{abstract}
	We predict frequency multiplication of surface acoustic waves in dielectric substrates via the ferromagnetic resonance of adjacent magnetic transducers when driven by microwaves. We find \textit{pure} second harmonic generation (SHG) without any linear and third harmonic components by a magnetic nanowire. The SHG and linear phonon pumping are switched by varying the saturated magnetization direction of the wire, or resolved directionally when pumped by magnetic nano-disc. We address the high efficiency of SHG with comparable magnitude to that of linear response, as well as unique non-reciprocal phonon transport that is remarkably distinct in different phonon harmonics. Such acoustic frequency comb driven by microwaves should bring unprecedented tunability for the miniaturized phononic and spintronic devices. 
	\end{abstract}
	
	\maketitle
	
\textit{Introduction}.---Surface acoustic waves (SAWs) are important information carriers in phononic and electronic devices \cite{classical_information_1,classical_information_2}, but also act as excellent information mediators for quantum communication in high-quality piezoelectric substrates  \cite{SAW_quantum_PRX,SAW_quantum_1,SAW_quantum_2}. Downscaled phononic devices rely on the generation of coherent phonons of above GHz frequency and sub-micrometer wavelength, which arguably represents one challenge by conventional electric approach so far since its excitation efficiency is low and energy consumption is high \cite{electric_1,electric_2,high_frequency_phononics}.
 In contrast, the ferromagnetic resonance (FMR) of magnetic nanostructures can pump such phonons via the magnetostriction efficiently in conventional dielectric substrates \cite{GGGmaterial,GGGattenuation,phononpumping1,GGGYIG,phonontransport,SAW_PRL,bright_dark_phonon,magnetphonon,phonondiode,Kei,PR_chirality}, which achieves efficient communication of spin information over millimeter distance.  The inverse process, i.e., the modulated transmission of SAWs via magnetostriction, was verified decades ago in a magnetoelastic bilayer towards SAW isolator functionality  \cite{Lewis_1972}, but recently obtains tremendous attention due to its remarkable non-reciprocity or diode effect with large on-off ratios observed in many ferromagnet$|$piezoelectric insulator heterostructures \cite{acoustic_pumping_1,acoustic_pumping_2,SAW_chiral_attenuation,SAW_chiral_attenuation,Onose_exp,Otani_exp,nonreciprocity_DMI,Page}. Most of these studies focus on the linear response, however, a regime limiting the tunability and maximal frequency of resonant phonons.

High phonon harmonics in the acoustic frequency multiplication, or the acoustic frequency comb, operate at a higher frequency and shorter wavelength than their linear component \cite{SHG_1,SHG_2,ultrasonic_1,nature_electronics}. In crystals their coherent generation relies on the anharmonic interaction of lattice and thus needs to exploit strong laser fields to achieve the demanding nonlinearity that may cause unavoidable parasitic effects such as heating and dephasing.  Second harmonic generation (SHG) of phonons in the terahertz frequency was excited in ultrashort time scales \cite{SHG_1,SHG_2}, where strong laser pulses are exerted, as well as in the megahertz frequency for the ultrasonic waves \cite{ultrasonic_1}. Without piezoelectricity \cite{nature_electronics}, achieving such nonlinearity for GHz phonons appears to be a formidable task.
Different from the electric approach, nonlinear magnetization dynamics for frequency multiplication is easily accessible, energy-saving, and well controlled by microwaves \cite{roadmap,PR_insulator,Bimu,Toeno_nonlinear,frequency_comb_Science,spin_wave_frequency_comb}.

\begin{figure}[htp!]
	    \centering
\includegraphics[width=0.95\textwidth]{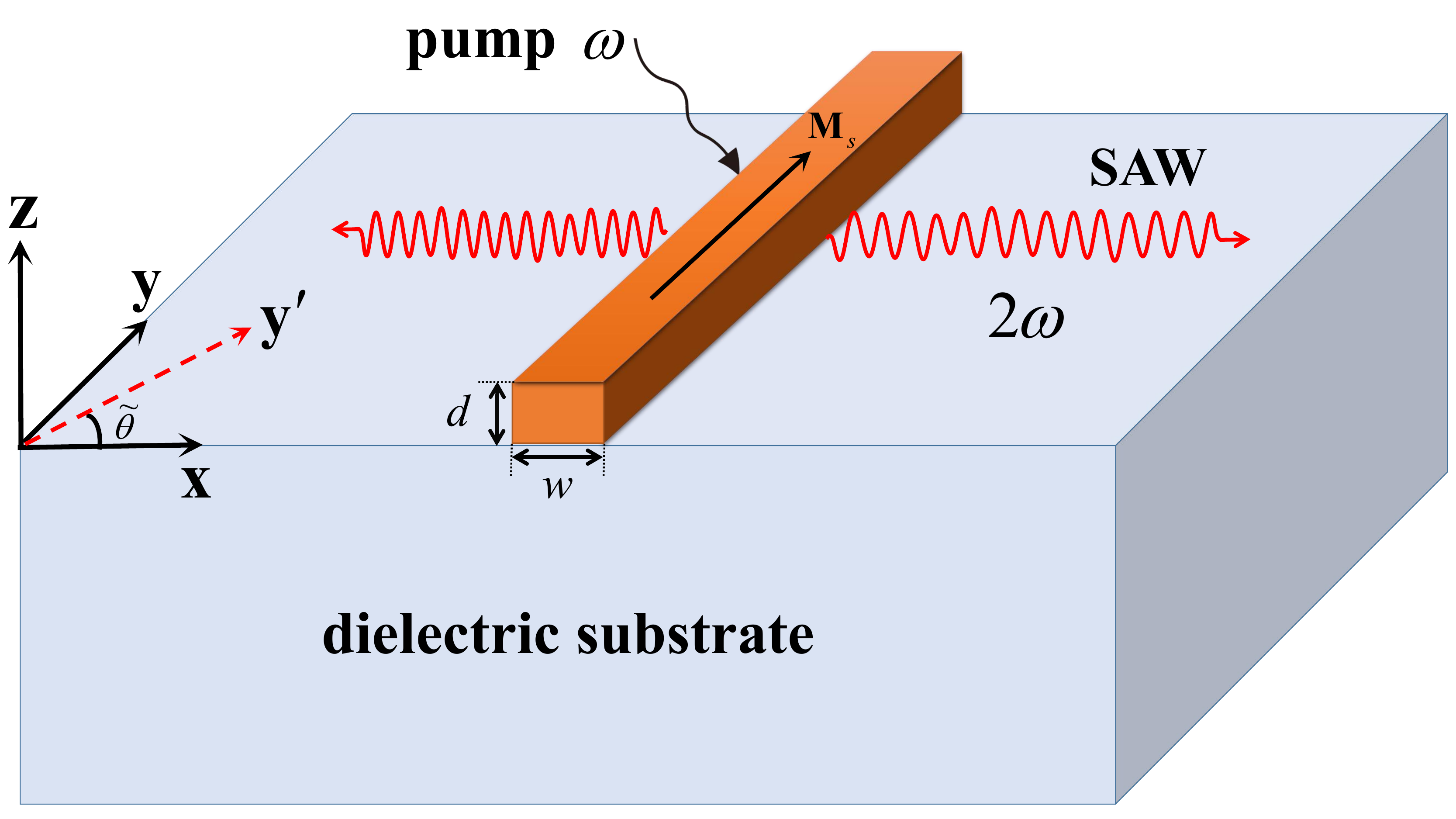}
	    \caption{Pure SHG $2\omega$ of SAWs in conventional dielectric substrates via phonon pumping by the FMR of a magnetic nanowire of thickness $d$ and width $w$ that is launched by microwaves of frequency $\omega$. The saturated magnetization ${\bf M}_s$ is biased by an external magnetic field, allowing the pure second harmonics of SAWs to mix with other components when away from the wire direction.}
	    \label{model}
\end{figure}

In this Letter, we predict the acoustic frequency multiplication as well as \textit{pure} SHG of SAWs of $\sim 10$~GHz frequency in conventional non-piezoelectric dielectric substrates, in which the linear and third harmonic harmonics completely vanish, via the phonon pumping of adjacent magnetic transducers that are launched by microwaves, as sketched in Fig.~\ref{model} for the magnetic nanowire configuration (such a wire is later replaced by a magnetic nano-disc). 
We can switch the pure SHG, when the saturated magnetization is along the wire direction, to the dominant linear phonon excitation, when the magnetization is biased to the wire normal direction, or realize their mixing flexibly with other arbitrary magnetization directions. All such phenomena can be exhibited conveniently with an in-plane magnetized nano-disc, where the pure SHG and linear response are resolved directionally. We find the efficiency of such SHG is high since with accessible magnetization nonlinearity its magnitude is not smaller than that of linear response, but the non-reciprocity appearing in the linear phonon pumping is strongly altered in the SHG due to the distinct dynamic magnetoelastic boundary conditions.

\textit{Model and acoustic frequency multiplication}.---The magnetoelastic heterostructure that we consider contains a nano-magnet ``M" of thickness $d$ with an in-plane equilibrium magnetization ${\bf M}_s$, such as a magnetic nanowire of width $w$ or a magnetic nano-disc of radius $r$, and an adjacent thick dielectric substrate ``N", which couple via the magnetostriction 
\cite{SAW_PRL,Kittel_old,parameters,Kei,nonreciprocity_DMI,phononpumping1}
\[F_{\rm me}=\frac{1}{M_s^{2}}\int d{\bf r} \left(B_{||} \sum_{i}  \varepsilon_{i i}M_{i}^{2}+B_{\perp} \sum_{i \neq j}  \varepsilon_{i j}M_{i} M_{j}\right),
	\]
	where $B_{||}$ and $B_{\perp}$ are the magneto-elastic coupling constants, $\{i,j\}=\{x,y,z\}$ denote the spatial index, and $\varepsilon_{i j}\equiv({1}/{2})\left({\partial u_{i}}/{\partial r_{j}}+{\partial u_{j}}/{\partial r_{i}}\right)$ is the strain tensor defined via the displacement field ${\bf u}$. Here we focus on the FMR of the nano-magnet \cite{GGGYIG,bright_dark_phonon}, such that the precessing magnetization ${\bf M}(t)$ can be treated as a macrospin, which is governed by the Landau-Lifshitz-Gilbert (LLG) equation \cite{Gilbert,Landau}:
\begin{equation}
    {\partial {\bf M}}/{\partial t}=-\mu_0\gamma {\bf M} \times {\bf H}_{\rm e f f}+\alpha ({\bf M}/{M_s}) \times {\partial {\bf M}}/{\partial t},
    \label{LLG}
\end{equation}	
where $\mu_0$ is the  vacuum permeability, $-\gamma$ is the
electron gyromagnetic ratio, and $\alpha$ is the phenomenological damping constant \cite{Gilbert}. The magnetization precesses around an effective magnetic field ${\bf H}_{\rm e f f}={\bf H}_{\rm app}+{\bf H}_d+{\bf H}_{\rm ex}+{\bf H}_e$  that contains the external field ${\bf H}_{\rm app}$ including the static ${\bf H}_0$ and dynamic ${\bf h}(t)$ fields, the demagnetizing field ${\bf H}_d=(-N_{xx}M_x,-N_{yy}M_y,-N_{zz}M_z)$, where $N_{x x}\simeq {d}/({d+w})$, $N_{y y}\simeq 0$,   and $N_{z z}\simeq {w}/({d+w})$ parameterize the demagnetization factor of the wire \cite{magnon_trap}, the exchange field ${\bf H}_{\rm ex}=A_{\rm ex}\nabla^2{\bf M}$ with the exchange stiffness $A_{\rm ex}$, as well as the effective field due to the magnetostriction 
\begin{align}
    H_{e,i}&\equiv-({1}/{\mu_0}){\delta F_{\rm me}}/{{\delta { M}_i({\bf r})}}\nonumber\\
&=-\frac{2}{\mu_{0}M_s} \sum_{j}\varepsilon_{ij} M_j \left[\delta_{ij}B_{||}+(1-\delta_{ij})B_{\perp}\right].
\end{align}

As a backaction, the magnetization also affects the static and dynamic strains of the elastic heterostructure. We have to distinguish  the displacement fields in the nano-magnet ${\bf u}_{\rm M}({\bf r},t)$ and the dielectric substrate ${\bf u}_{\rm N}({\bf r},t)$, as well as their different material densities $\rho_{\rm M}$ and $\rho_{\rm N}$ in the elastic equations of motion \cite{BCs,phonondiode,elastic1,elastic2}:
\begin{align}
\rho_{\rm M} \ddot{\bf {u}}_{\rm M}&=\nabla\cdot ( \overleftrightarrow{{\pmb{\sigma}}_{\rm M}}+\overleftrightarrow{\pmb  {\mathcal \eta}}),\nonumber\\
\rho_{\rm N} \ddot{\bf {u}}_{\rm N}&=\nabla \cdot \overleftrightarrow{{{\pmb\sigma}}_{\rm N}},
\label{EOM_u}
	\end{align}
which are governed by the mechanical stress tensor
\begin{align}
    \sigma^{\rm N,M}_{ij}=\delta_{ij}\lambda_{\rm N,M}\sum_l\varepsilon^{\rm N,M}_{ll}+2\mu_{\rm N,M}\varepsilon^{\rm N,M}_{ij},
\end{align}
where $\lambda_{\rm N,M}$ and $\mu_{\rm N,M}$ are the associated Lam\'{e} constants,
and the magnetization stress tensor inside the magnet
\begin{align}
 \eta_{ij}&\equiv \partial F_{\rm me}/\partial(\partial u_i/\partial r_j)\nonumber\\
&= M_i M_j [\delta_{ij}B_{||}+(1-\delta_{ij})B_{\perp}]/M_s^2.
\label{magnetization_stress}
\end{align} 
However, with uniform magnetization there is no net effect of $\overleftrightarrow{\pmb  {\mathcal \eta}}$ inside the magnet since $\nabla \cdot \overleftrightarrow{\pmb  {\mathcal \eta}}=0$.
All the phonon pumping effect thereby comes from the static and dynamic magnetization stress at the boundary of the magnet, which appears in the boundary conditions defined by the continuity of the force per unit area or the stress vector at the surfaces and interfaces \cite{BCs,phonondiode,elastic1,elastic2,Kei}: 
\begin{align}
\overleftrightarrow{\pmb{\sigma}_{\rm N}}\cdot{\bf n}|_{\rm A}&=0,\nonumber\\
(\overleftrightarrow{\pmb{\sigma}_{\rm M}}+\overleftrightarrow{\pmb {\mathcal \eta}})\cdot{\bf n}|_{\rm B}&=0,\nonumber\\
(\overleftrightarrow{{\pmb {\mathcal \sigma}}_{\rm M}}+\overleftrightarrow{\pmb {\mathcal \eta}})\cdot{\bf n}|_{\rm C}&=\overleftrightarrow{{\pmb {\mathcal\sigma}}_{\rm N}}\cdot{\bf n}|_{\rm C}.
\label{boundary_conditions}
	\end{align}
Here we denote the interfaces between the dielectric substrate and vacuum as ``A", between the nano-magnet and vacuum as ``B", and between the dielectric substrate and nano-magnet as ``C", such that $\bf n$ is the normal unit vector of each interface.

When ${\bf M}_s$ is aligned to the wire $\hat{\bf y}$-direction, the  dynamic boundary magnetization stress in the linear order of fluctuated magnetization $\overleftrightarrow{\pmb {\mathcal \eta}}\cdot{\bf n}|_{\rm C}=M_zB_{\perp}/M_s\hat{\bf y}$ is along the wire direction, as well as $(\overleftrightarrow{\pmb {\mathcal \eta}}\cdot{\bf n}|_{\rm B})\parallel\hat{\bf y}$, which thereby excites no SAW propagating \textit{normal} to the wire direction since its associated mechanical stress vector along the wire direction vanishes and thereby mismatches. We thereby expect the absence of linear harmonics for the pumped SAWs propagating normally to the wire in such a magnetic configuration.

We substantiate such expectation by numerical simulations, but allow an arbitrary in-plane saturated magnetization by an angle $\tilde{\theta}$ with respect to the wire normal $\hat{\bf x}$-direction (Fig.~\ref{model}), which we find is nearly parallel to the external static field ${\bf H}_0$ \cite{supplement}. We combine the LLG equation (\ref{LLG}) with the elastic equations of motion (\ref{EOM_u}) under the boundary conditions (\ref{boundary_conditions}) in COMSOL Multiphysics \cite{COMSOL, COMSOLblog}. We choose the nano-magnet as the yttrium iron garnet (YIG) nanowire  \cite{YIG_nanowire} of thickness $d=80$~nm, width $w=150$~nm, and saturated magnetization $\mu_0 M_s=0.177$~T \cite{GGGYIG,bright_dark_phonon},  biased by $\mu_0H_0=0.1$~T, which has high magnetic quality with $\alpha=10^{-4}$. It is adjacent to a thick gadolinium gallium garnet (GGG) substrate that has high acoustic quality \cite{GGGYIG,bright_dark_phonon}. Their elastic properties are close but not identical: for YIG, $\rho_{\rm M}=5170$~${\rm kg/m^3}$, $\lambda_{\rm M}=1.16\times10^{11}$~${\rm N/m^2}$, $\mu_{\rm M}=7.64\times10^{10}$~${\rm N/m^2}$  \cite{phononpumping1}, while for GGG, $\rho_{\rm N}=7080$~${\rm kg/m^3}$, $\lambda_{\rm N}=1.27\times10^{11}$~${\rm N/m^2}$, $\mu_{\rm N}=8.83\times10^{10}$~${\rm N/m^2}$  \cite{GGG}. They are coupled via magnetostriction with the coupling constants $B_{||}=3.48\times10^5$~${\rm J/m^3}$ and $B_{\perp}=6.96\times10^5$~${\rm J/m^3}$  \cite{phononpumping1}. The sound velocity of SAWs  $c_r=3271.8$~${\rm m/s}$ \cite{Viktorov1967}.

We apply an in-plane broadband magnetic field transverse to the saturated magnetization ${\bf h}(t)=h_0\sin(\omega_{\rm F} t)\hat{\bf x}'$ with a short duration time $0\le t\le 2\pi/\omega_{\rm F}$, where $\hat{\bf x}'\perp {\bf M_s}\perp \hat{\bf z}$, with which we adjust the FMR frequency $\omega_{\rm F}/(2\pi)=\{5.43,3.71,2.29\}$~GHz and the field strength $\mu_0h_0=\{9.05,6.20,3.83\}$~mT to make sure the pumped transverse magnetization $M_{z'}\approx 0.15M_s$ or the precession angle $\sim 8.5^{\circ}$.

\begin{figure}[htp]
\centering
\hspace{-0.45cm}
\includegraphics[width=0.36\textwidth,trim=0.1cm 1cm 0.2cm 0.2cm]{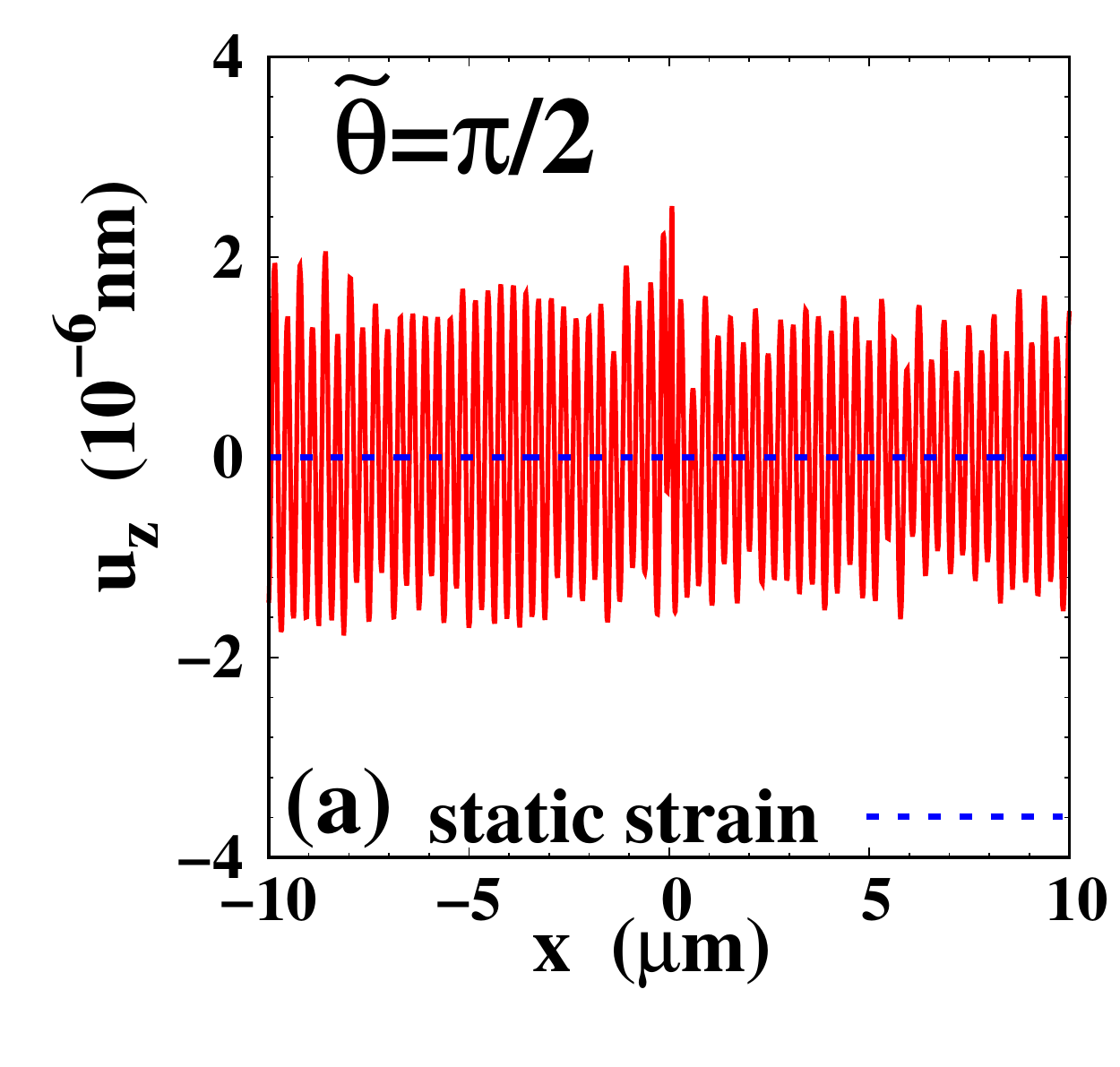}
\hspace{-0.38cm}
\includegraphics[width=0.36\textwidth,trim=0.1cm 1cm 0.2cm 0.2cm]{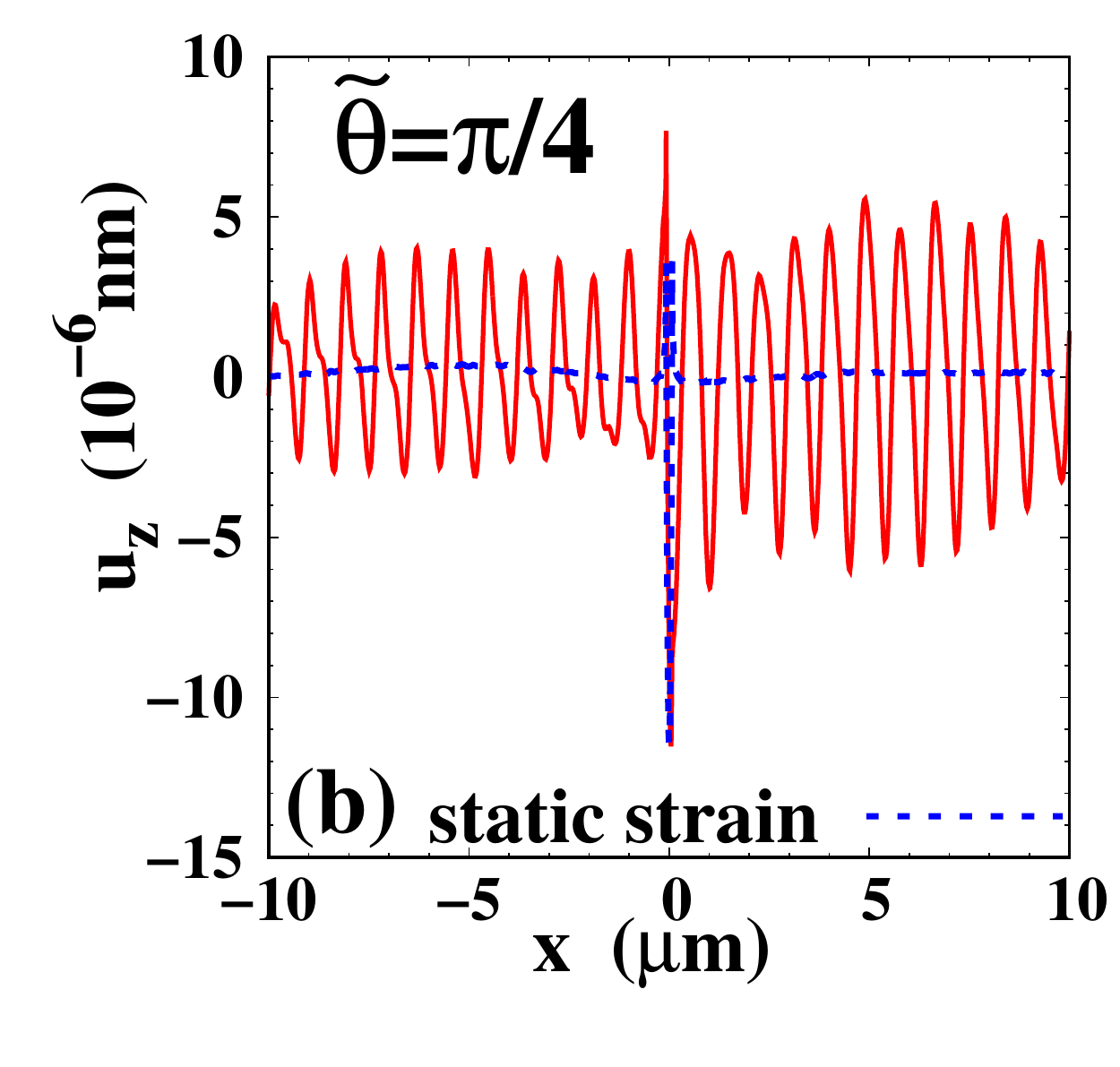}
\hspace{-0.35cm}
\includegraphics[width=0.36\textwidth,trim=0.1cm 1cm 0.2cm 0.2cm]{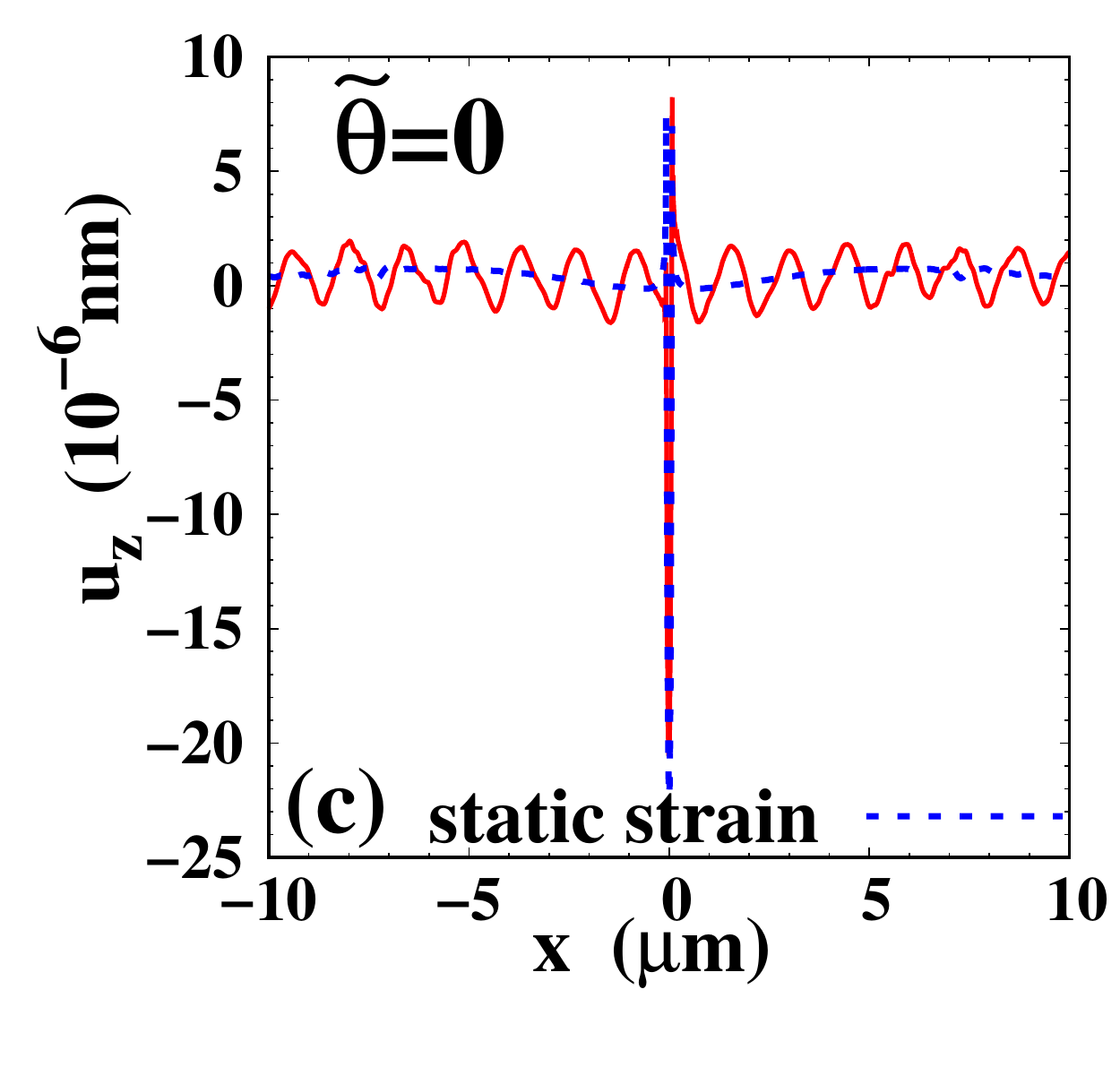}\\
\hspace{-0.45cm}
\includegraphics[width=0.36\textwidth,trim=0.1cm 0.5cm 0.2cm 0.1cm]{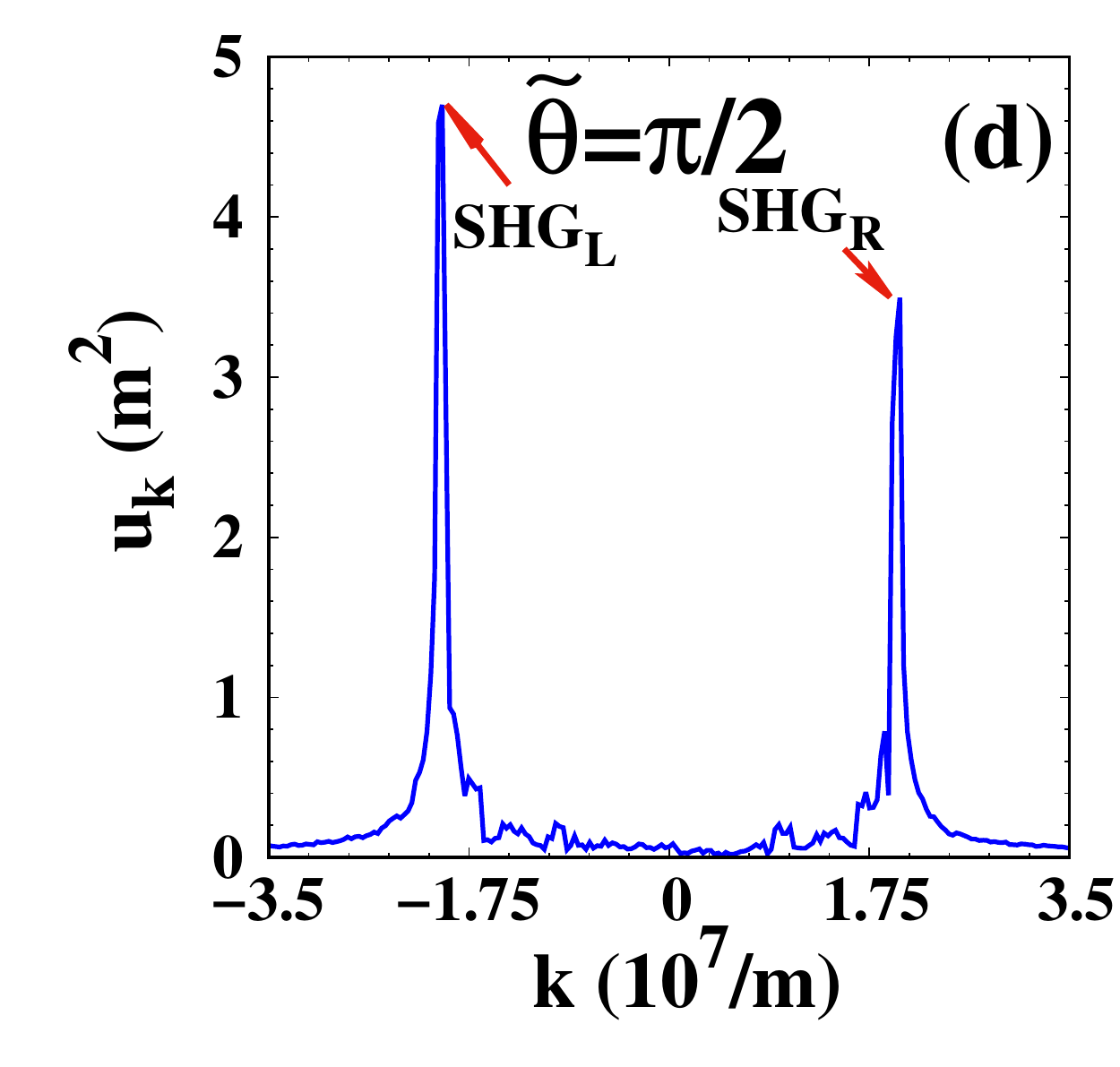}
\hspace{-0.38cm}
\includegraphics[width=0.36\textwidth,trim=0.1cm 0.5cm 0.2cm 0.1cm]{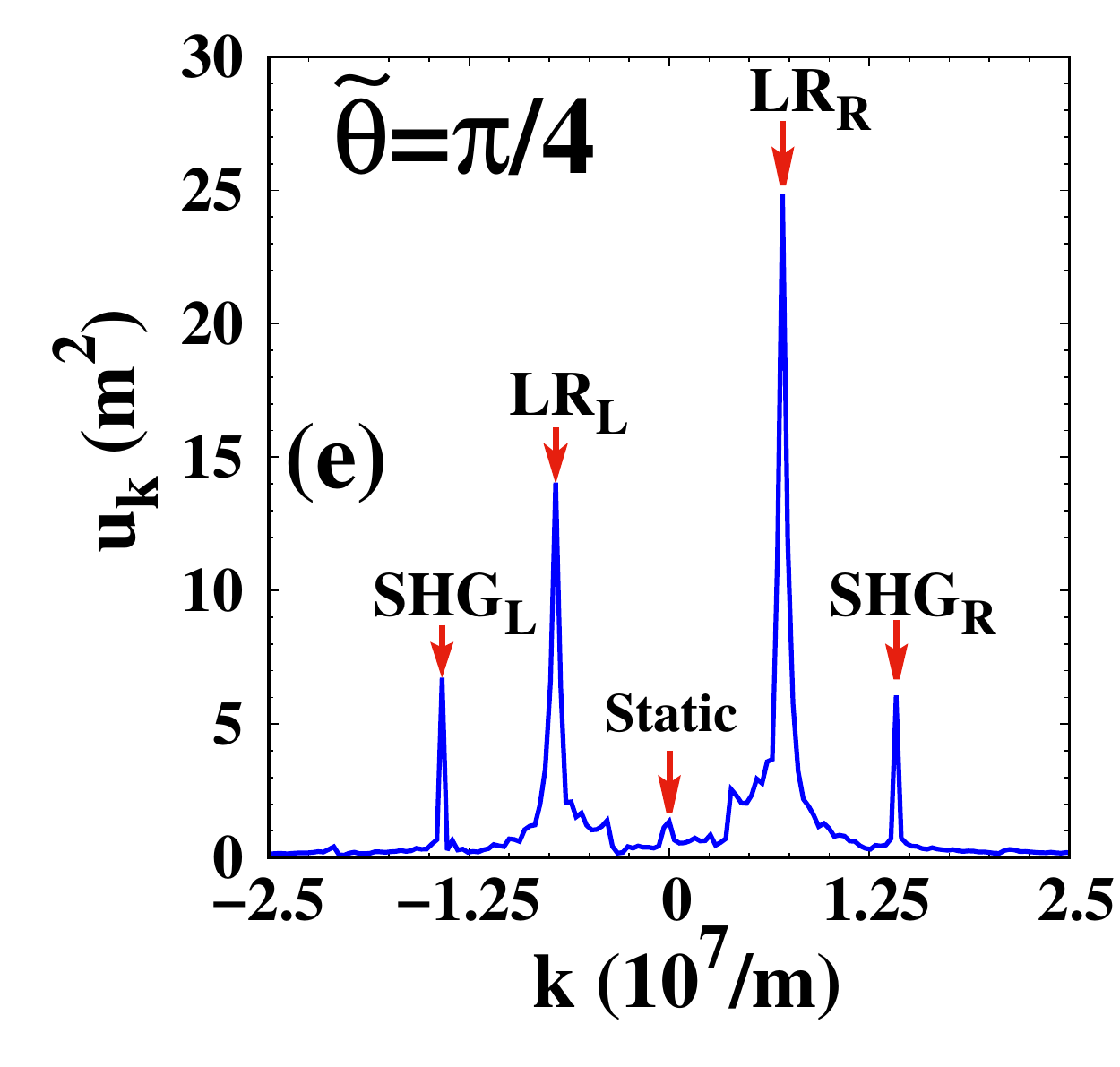}
\hspace{-0.35cm}
\includegraphics[width=0.36\textwidth,trim=0.1cm 0.5cm 0.2cm 0.1cm]{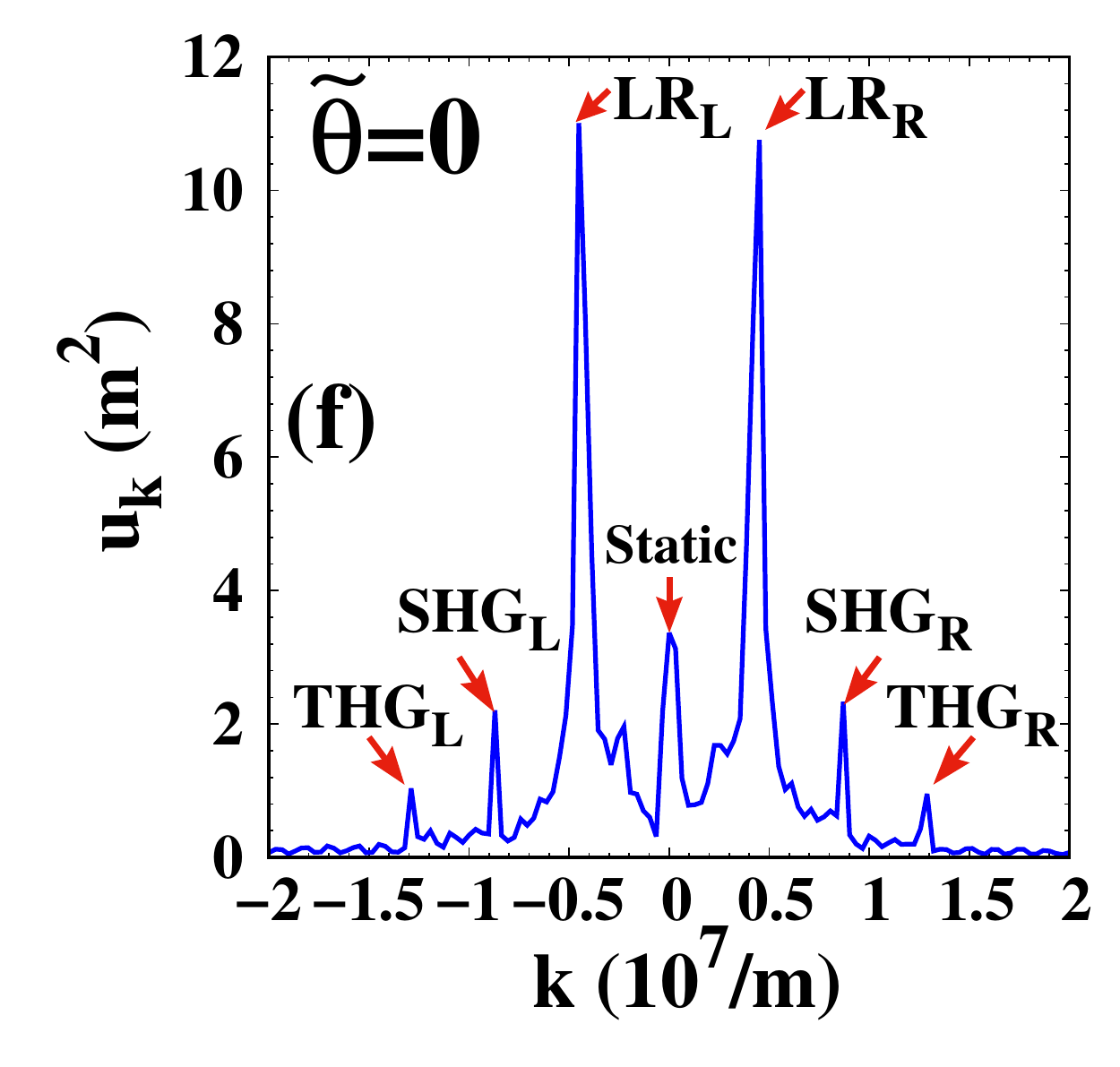}\\
\hspace{-0.45cm}
\includegraphics[width=0.36\textwidth,trim=0.1cm 0.5cm 0.2cm 0.1cm]{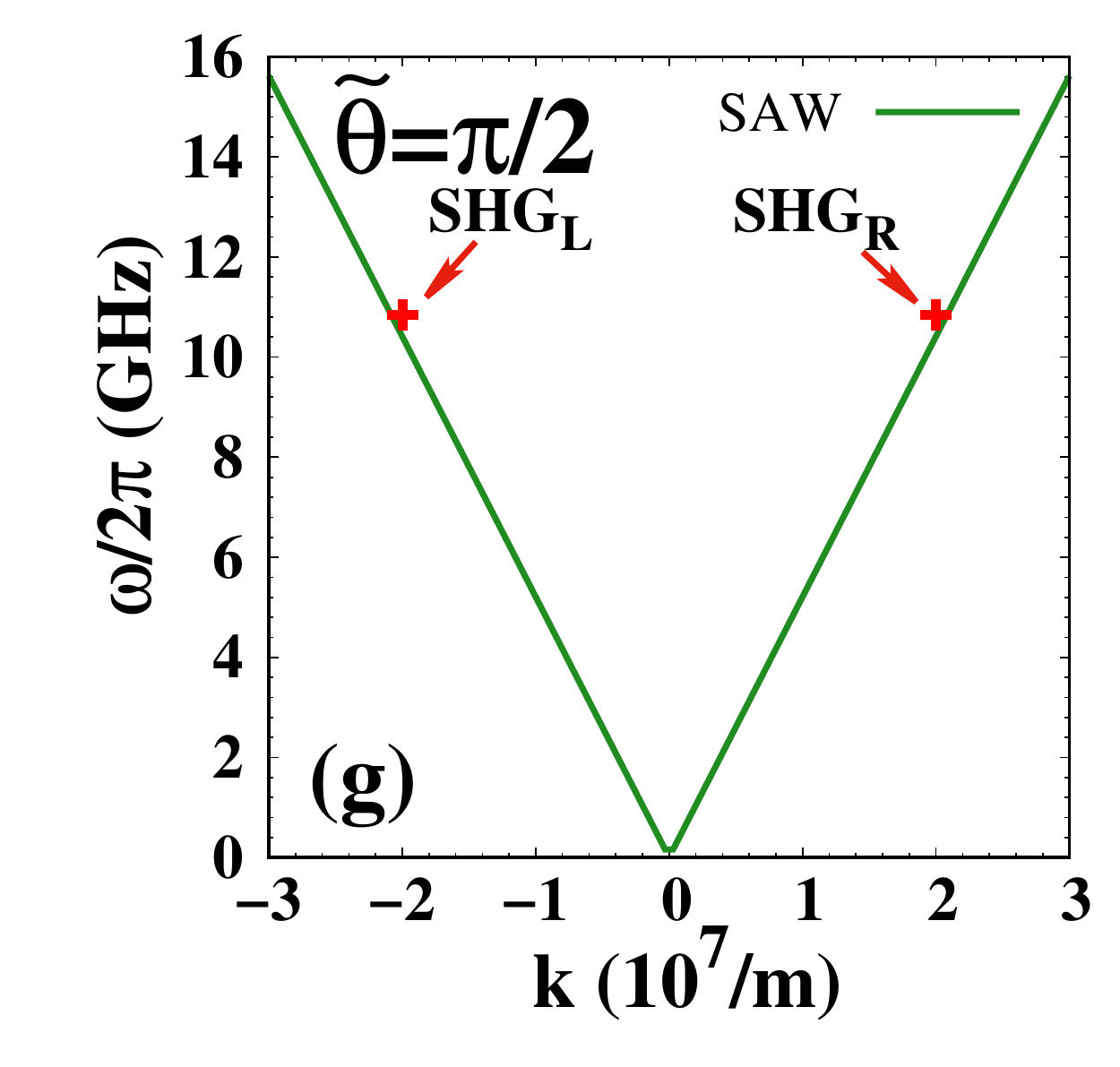}
\hspace{-0.38cm}
\includegraphics[width=0.36\textwidth,trim=0.1cm 0.5cm 0.2cm 0.1cm]{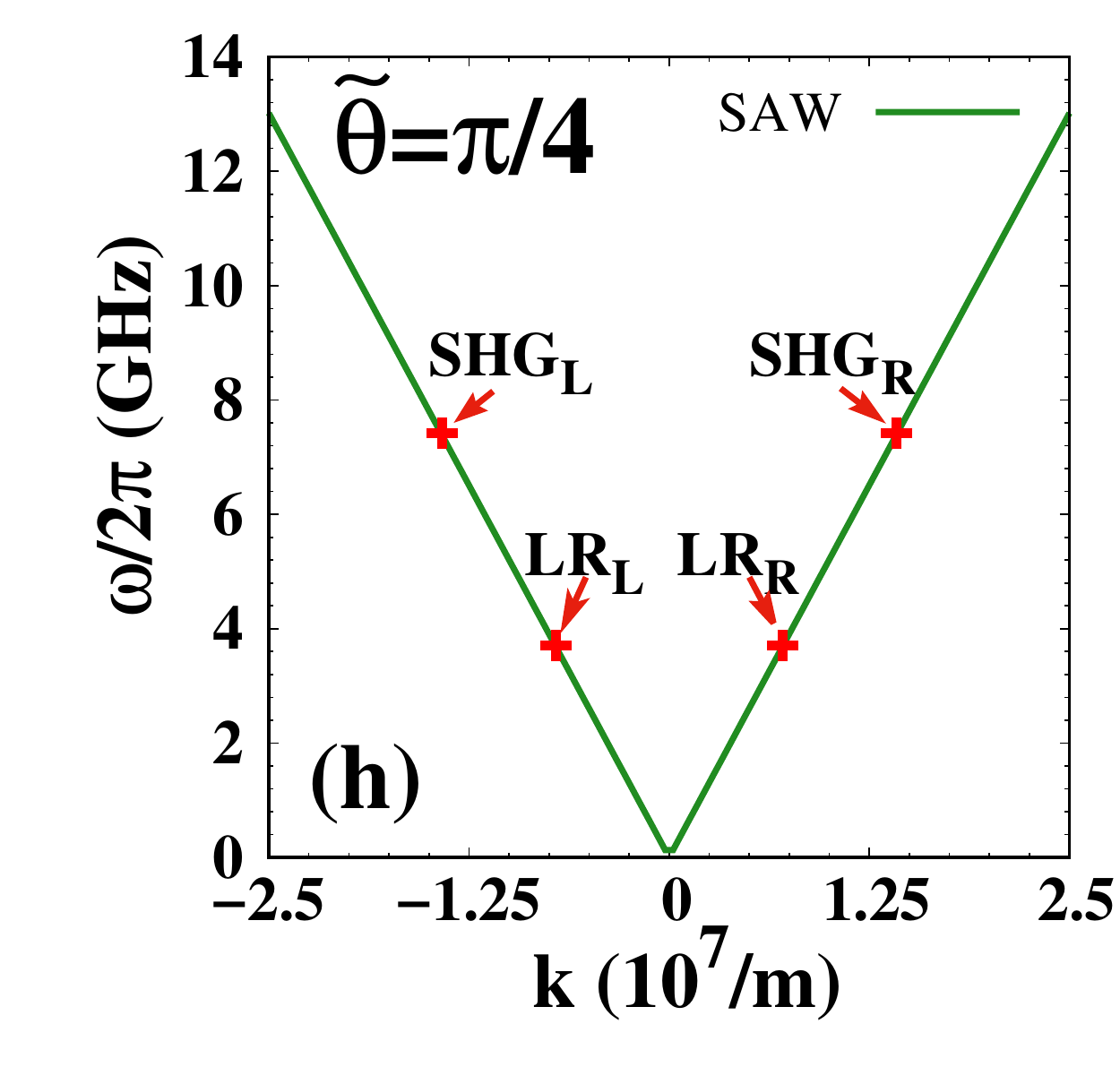}
\hspace{-0.35cm}
\includegraphics[width=0.36\textwidth,trim=0.1cm 0.5cm 0.2cm 0.1cm]{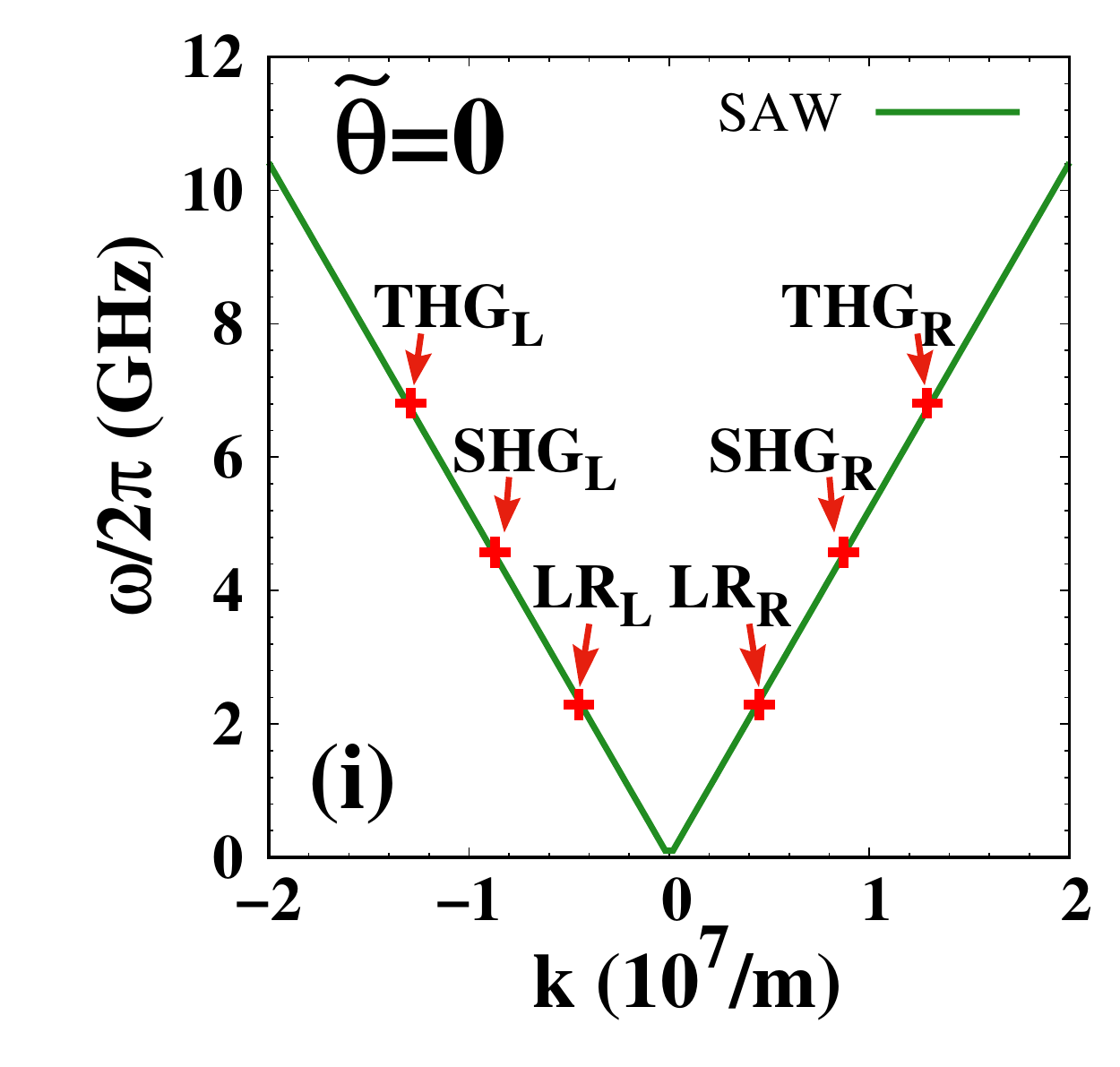}
 \caption{Acoustic frequency multiplication in GGG substrates by the FMR of adjacent YIG nanowire with different magnetic configurations, plotted about $10$~ns after the end of magnetic-field pulse. The blue dashed lines in (a-c) indicate the static strain that is pronounced only near the nano-magnet. When ${\bf M}_s$ is aligned to the wire direction with $\tilde{\theta}=\pi/2$, there is only the SHG in the pumped SAWs, as shown by the displacement field $u_z$ at the surface $z=0$ in (a), the resolved Fourier component $u_k$ in (d), as well as the oscillation frequency and wave vector of $u_z$ by crosses in comparison with the SAW dispersion in (g). In the other magnetic configurations with $\tilde{\theta}=\pi/4$ [(b), (e), (h)] and $\tilde{\theta}=0$ [(c), (f), (i)], the LR, SHG, and THG coexist with flexible tunability by magnetization directions.}
\label{simulation_results}
\end{figure}

Figure~\ref{simulation_results} plots the frequency multiplication of SAWs up to the third harmonic generation (THG) by the FMR of YIG nanowire with different magnetic configurations, characterized by the pumped displacement field $u_z$ at the surface $z=0$ (a-c), their Fourier components $u_k$ (d-f), as well as the oscillation frequency and wave number resolved from the peaks in $u_k$ in comparison to the SAW dispersion (g-i). The excellent agreement of the oscillation frequency and wave vector with the SAW dispersion $\omega_k=c_r|k|$ implies that the pumped elastic strain at the surface is dominated by SAWs. Static strains exist only near the nano-magnet but vanish when ${\bf M}_s$ is along the wire direction because of the absence of static $\overleftrightarrow{\pmb  {\mathcal \eta}}$.

One remarkable feature in Fig.~\ref{simulation_results} is that when the saturated magnetization is along the wire direction ($\tilde{\theta}=\pi/2$), there is only the SHG of SAWs propagating normally to the wire, without any linear and third harmonics, while when ${\bf M}_s$ is normal to the wire direction ($\tilde{\theta}=0$), the linear response (LR) dominates. Such SHG comes completely from the nonlinearity of magnetic stress at the boundary, which scales as $h_0^2$ in its amplitude, thus distinguished from the 
anharmonicity effect of lattice \cite{SHG_1,SHG_2}. The mixing of SHG and other harmonics is realized when ${\bf M}_s$ is away from the parallel setup, e.g., $\tilde{\theta}=\pi/4$ in Fig.~\ref{simulation_results}(b,e,h). The THG is unique since it comes from the interaction between magnons but not the nonlinear magnetic stress $\propto \exp({2i \omega_{\rm F}t})$ \cite{supplement}. These provide flexible tunability for the demanding phonon frequency achievable by different directions and magnitudes of the static magnetic field.

Pronounced non-reciprocity exists in the linear phonon pumping, as shown in Fig.~\ref{simulation_results}(e) when $\tilde{\theta}=\pi/4$. Such non-reciprocity vanishes when the magnetization is normal to the wire direction as in Fig.~\ref{simulation_results}(f). These numerical results agree with the theoretical expectations from the previous analytical solutions \cite{SAW_PRL,phonondiode}. However, in the SHG the non-reciprocity is generally suppressed in almost all the magnetic configurations, as shown by the Fourier components with opposite momenta in Fig.~\ref{simulation_results}(d,e,f).

\textit{Quantum formalism.}---Above simulated  phonon high harmonic generation can best be formulated in a quantum language. The magnetization operators 
$\hat{{M}}_{x', z'}\simeq -\sqrt{2 \gamma \hbar M_{s}}({\cal M}_{x',z'} \hat{m}+{\cal M}_{x', z'}^{*} \hat{m}^{\dagger})+{\cal O}(\hat{m}^3)$ and $\hat{M}_{y^{\prime}}\simeq M_s-\gamma\hbar\left[{\cal M}_{z^{\prime}}^2({\bf r})+{\cal M}_{x^{\prime}}^2({\bf r})\right] \hat{m} \hat{m}-\gamma\hbar\left[{\cal M}_{z^{\prime} }^{* 2}({\bf r})+{\cal M}_{x^{\prime}}^{* 2}({\bf r})\right] \hat{m}^{\dagger} \hat{m}^{\dagger}+{\cal O}(\hat{m}^3)$ contain the linear Holstein-Primakoff \cite{HP,Kittel_book,Walker_sphere,magnetic_nanodots,Sanchar_PRB,surface_roughness} expansion of Kittel magnon $\hat{m}$ and their dominant interactions with strength governed by the ellipticity of eigenmodes ${\cal M}_{x'}=i \xi_m^2{\cal M}_{z'}$ \cite{supplement}, which are not circularly polarized when the form factor $\xi_m\ne 1$.  The eigenmodes ${\pmb {\cal U}}(x, z, k)$ of SAWs in the elastic heterostructure contain both near-field solution close to the magnet and far-field $|x|\gg w/2$ solution  that converges asymptotically to those of SAWs \cite{Viktorov1967}. In terms of them and the SAW operator $\hat{p}_{k}$, we quantize the displacement field
	\begin{align}
	    \hat{\mathbf{u}}(x, z, t)=\sum_{k}\left({\pmb {\cal U}}(x, z, k) \hat{p}_{k}+{\pmb {\cal U}}^{*}(x, z, k) \hat{p}_{k}^{\dagger}\right).
	    \label{displacement_field}
	\end{align}
	 Substituting into magnetostriction energy, we obtain the magnon-phonon coupling Hamiltonian
    $\hat{H}_{c}=\hbar \sum_{n\ge 1}\sum_{k} g^{(n)}_{k} (\hat{m}^{\dagger})^n \hat{p}_{k}+{\rm H.c.}$ (refer to the Supplementary Material \cite{supplement} for details), 
where the $n$-th order coupling constants $g^{(n)}_k$ rely on the near-field solution of SAWs. $g_k^{(3)}$ vanishes for the circular precession $\xi_m=1$, and when ${\bf M}_s$ is parallel to the wire direction, $g_k^{(1)}$ and $g_k^{(3)}$ vanish, leading to the pure SHG \cite{supplement}. So the linear fluctuation of $\hat{m}$ is responsible for the LR and SHG of SAWs, while the double frequency in $\hat{m}^2$, existing in elliptical precessions $\xi_m\ne 1$, causes the THG. The interaction is ``non-reciprocal" when $|g^{(n)}_k|\ne |g^{(n)}_{-k}|$.

Including broadband microwaves ${\bf h}(t)=h_{x'}(t)\hat{\bf x}'$ and the damping rates of magnons and phonons $\delta_m$ and $\delta_p$, the magnon and surface phonon obey the Langevin's equations \cite{Langevin_1,Langevin_2}
\begin{align}
	d \hat{m} / d t &=-i (\omega_{\rm F}-i\delta_m) \hat{m}-i \sum_{n\ge 1}\sum_{k}ng^{(n)}_{k} (\hat{m}^{\dagger})^{(n-1)} \hat{p}_{k}\nonumber\\
	& -{\mu_0} \sqrt{{\gamma  M_s V}/({2\hbar})}\xi_m h_{x'}(t),\nonumber\\
	d \hat{p}_{k} / d t &=-i (\omega_{k}-i\delta_{p}) \hat{p}_{k}-i\sum_{n\ge 1}g_k^{(n)*} \hat{m}^n, 
	\label{Langevin_equation}
\end{align}
where $V=wdl$ is the wire's volume with length $l$. Here we focus on a large coherent pumping such that the  magnon's thermal population is much smaller than that driven by microwaves. To solve the nonlinear Eq.~(\ref{Langevin_equation}), we apply the mean-field approximation $\hat{A}\hat{B}=\langle \hat{A}\rangle \hat{B}+\hat{A}\langle\hat{B}\rangle$ for operators. Below we denote the ensemble-averaged $\langle\hat{A}\rangle=A$. Disregarding the far-off-resonant excitation, we find in the frequency domain the coherent amplitudes of SAWs
\begin{align}
	p_{k}(\omega)=G_k(\omega)\sum_{n\ge 1}\int dt_1e^{i\omega t_1}g_k^{(n)*}{m}^n(t_1)
	\label{p_k}
	\end{align}
contain all the harmonics of coherent magnon amplitude, where the phonon's Green function $G_k(\omega)={1}/({\omega-\omega_{k}+i \delta_{p}})$.
	
The magnon amplitude, on the other hand, should be self-consistently solved by the nonlinear equation, to the leading two orders of the coupling constants, 
		\begin{align}
		&{m}{(\omega)}=\frac{1}{\omega-\omega_{\rm F}+i \delta_{m}}\Big[\sum_{k} G_{k}(\omega)|g^{(1)}_{k}|^{2}{m}{(\omega)}
		\nonumber\\
		&+6\sum_{k, \omega_{1}, \omega_{2}} G_k(\omega+\omega_1)|g^{(2)}_{k}|^{2} {m}^*{\left(\omega_{2}\right)} {m}{\left(\omega_{1}\right)}{m}{\left(\omega+\omega_{2}-\omega_{1}\right)}\nonumber\\ 
		&-i{\mu_0} \sqrt{{\gamma  M_s V}/({2\hbar})}\xi_m h_{x'}(\omega)\Big].
		\label{magnon_equation_2}
		\end{align}	
Treating the phonon's back action to the FMR as a perturbation, here we pursue an iteration solution of Eq.~(\ref{magnon_equation_2}). Substituting the unperturbed solution ${m}^{(0)}(\omega)\approx
	-i{\mu_0} \sqrt{{\gamma  M_s V}/{2\hbar}}\xi_m{h_{x'}(\omega)}/({\omega-\omega_{\rm F}+i\delta_m})$ into (\ref{magnon_equation_2}), we arrive at  
\begin{align}
	{m}{(\omega)}\approx\frac{-i{\mu_0}\sqrt{{\gamma  M_s V}/({2\hbar})}\xi_m h_{x'}(\omega)}{\omega-\omega_{\rm F}+i \delta_{m}+\Sigma_{\rm L}(\omega)+\Sigma_{\rm NL}(\omega) },
	\label{m_omega}
		\end{align}
	where $\Sigma_{\rm L}(\omega)=-\sum_{k} G_k (\omega)|g^{(1)}_{k}|^{2}$ and $\Sigma_{\rm NL}(\omega)=-6l{n}_{m }\sum_{k} G_k (\omega+\omega_{\rm F})|g^{(2)}_{k}|^{2}$ are self energies contributed, respectively, by the linear and nonlinear phonon pumping. $n_m\equiv \langle\hat{m}^{\dagger}(t)\hat{m}(t)\rangle=[\mu_0\sqrt{{\gamma M_s wd}/{(2\hbar)}}\xi_mh_{x'}(\omega_\text{F})]^2$ is the pumped magnon number per unit wire length. Around FMR, the imaginary part of $\Sigma$, i.e., 
$\operatorname{Im}\Sigma_{\rm L}(\omega_{\rm F})={L}/({2c_r})(|g^{(1)}_{-k_r}|^2+\left|g^{(1)}_{k_r}\right|^2)$ and $\operatorname{Im}\Sigma_{\rm NL}(\omega_{\rm F})={3Ll{n}_{m}}/{c_r}(|g^{(2)}_{-2k_r}|^2+\left|g^{(2)}_{2k_r}\right|^2)$ contribute to linear and nonlinear magnon dampings, where $L$ is the substrate's length and $k_r={\omega_{\rm F}}/{c_r}$.

Substituting the solutions $p_k(t)$ (\ref{p_k}) and $m(t)$ (\ref{m_omega}) to the displacement field (\ref{displacement_field}), we close the momentum integral in the upper (lower) half complex plane when $x>0$ ($x<0$). When  $x>0$, the displacement field  
\begin{align}
		{\bf u}_R=\frac{2L}{c_r}\sum_{n\ge 1}{\rm Im} \left({\pmb {\cal U}}(z, n k_r)e^{i n k_rx}g^{(n)*}_{n k_r}{m}^n(t)\right),
		\label{right}
	\end{align}
	only appears on the right-hand side of the nanowire but on its left-hand side
	\begin{align}
		{\bf u}_L=\frac{2L}{c_r}\sum_{n\ge 1}{\rm Im} \left({\pmb {\cal U}}(z, -n k_r)e^{-i n k_rx}g^{(n)*}_{-n k_r}{m}^n(t)\right).
		\label{left}
	\end{align}
	Solutions (\ref{right}) and (\ref{left}) contain both the linear and nonlinear phonon pumping effects.
$|{\bf u}_L|\ne |{\bf u}_R|$ with the non-reciprocal couplings.

\begin{figure}[htp]
\centering
	\hspace{-0.27cm}\includegraphics[width=0.53\textwidth,trim=0.4cm 1cm 0.15cm 0.2cm,clip]{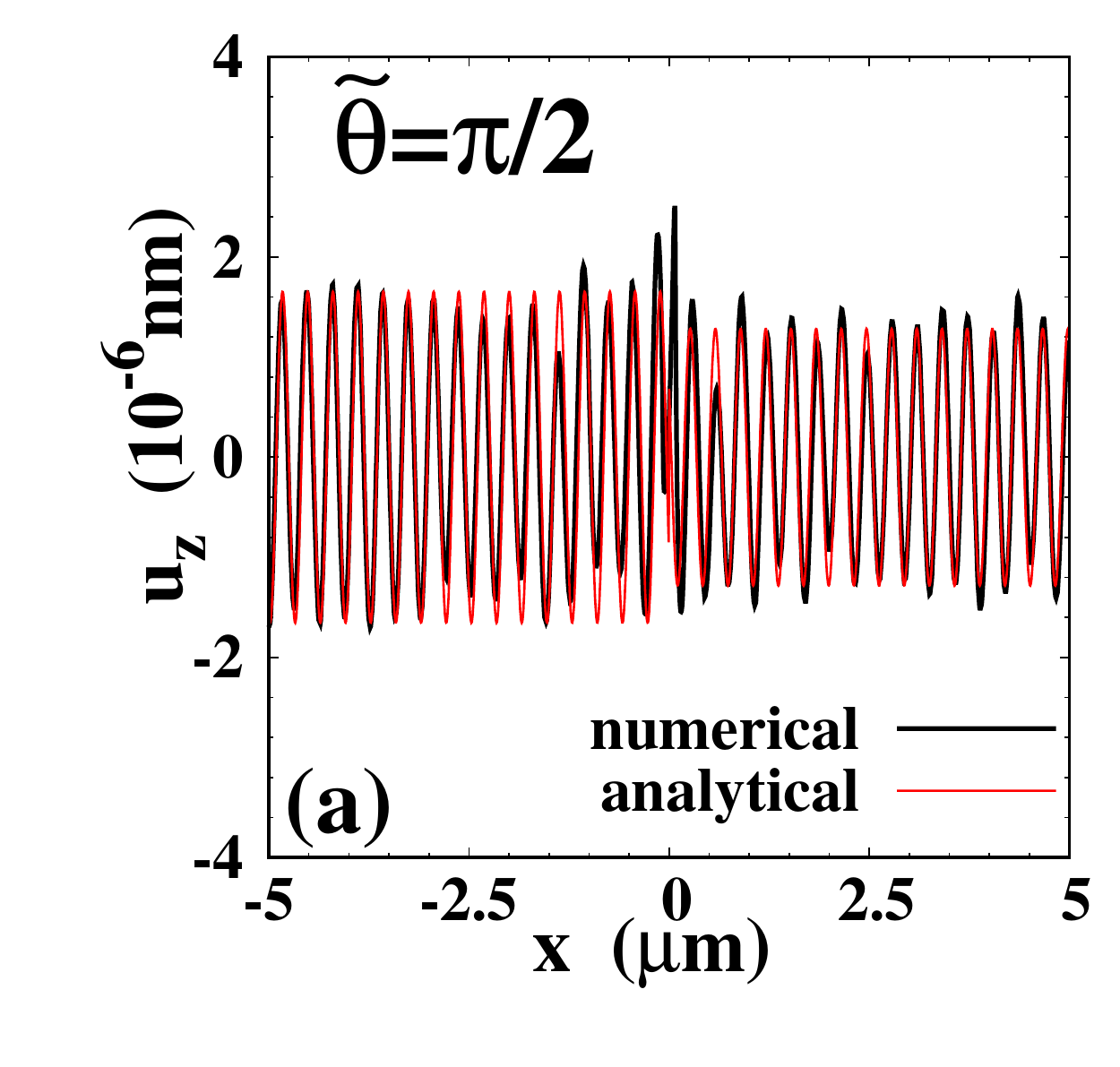}
	\hspace{-0.35cm}\includegraphics[width=0.53\textwidth,trim=0.4cm 1cm 0.15cm 0.2cm,clip]{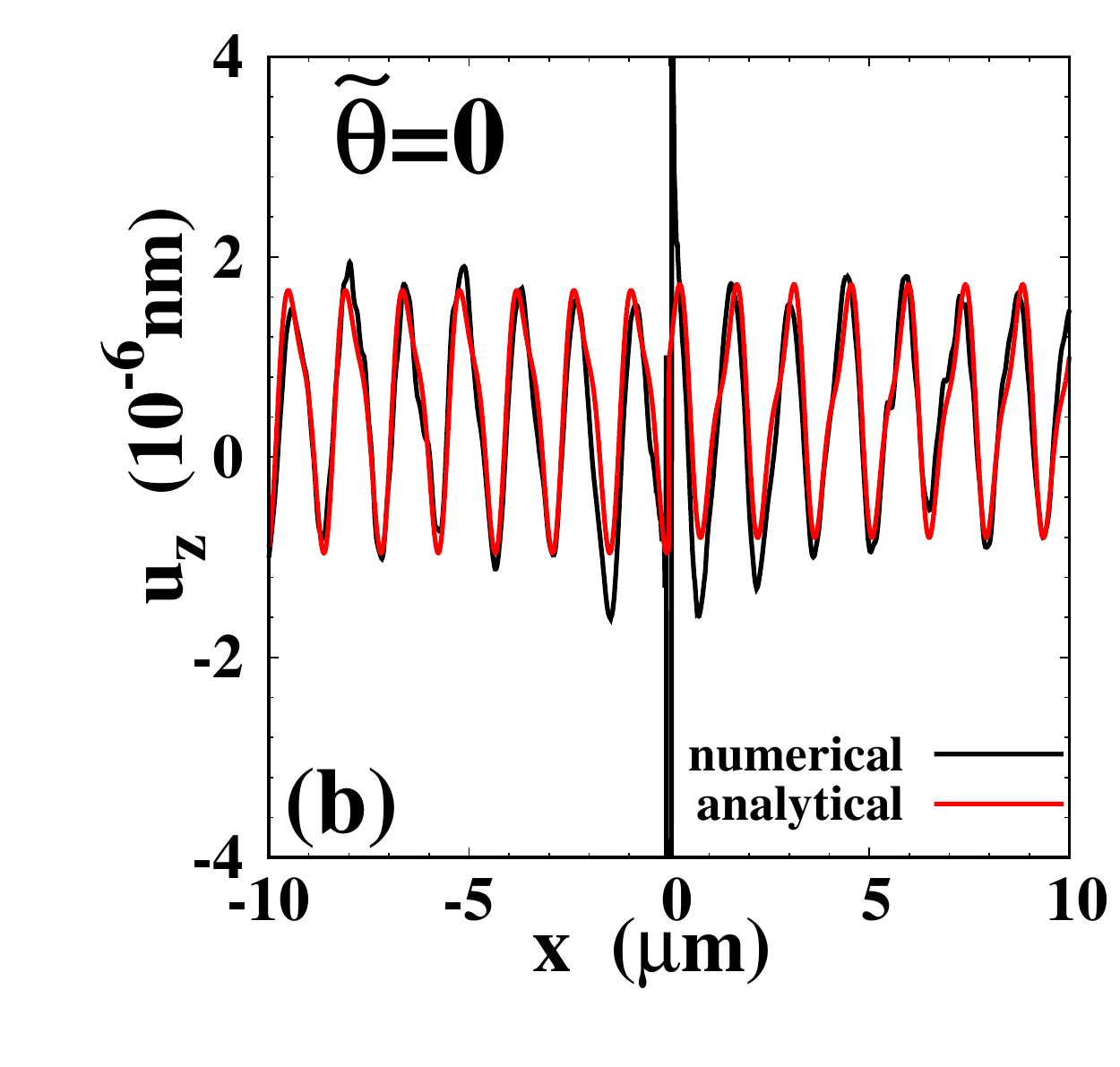}
\caption{Calculated SHG and linear phonon pumping by analytical solutions (\ref{right}) and (\ref{left}) with the simulation parameters. We use   $g^{(2)}_{2k_r}=0.76$~mHz and $g^{(2)}_{-2k_r}=0.98$~mHz in (a), and $g^{(1)}_{k_r}=1.2$~kHz, $g^{(1)}_{-k_r}=1.3$~kHz, $g^{(2)}_{2k_r}=0.21$~mHz, and $g^{(2)}_{-2k_r}=0.2$~mHz in (b).}
\label{analytical_solution}
\end{figure}

With the parameters in the simulation, solutions (\ref{right}) and (\ref{left}) reproduce the numerical results well with ${m}(t)\rightarrow \sqrt{{ldw}/({2\gamma\hbar M_s})}\left({i}M_{x'}(t)/{\xi_m}-\xi_m M_{z'}(t) \right)$, calculated from Eq.~(\ref{m_omega}) by disregarding the small damping, and proper coupling constants, as shown in Fig.~\ref{analytical_solution}. Our quantum formalism is thereby established for the future study of quantum communication with on-chip magnons \cite{Zouji,PRX_quantum,Bowen} mediated by high-quantity acoustic oscillation.

\begin{figure}[htp]
\centering
\includegraphics[width=0.495\textwidth,trim=0.1cm 0.1cm 0.15cm 0.5cm]{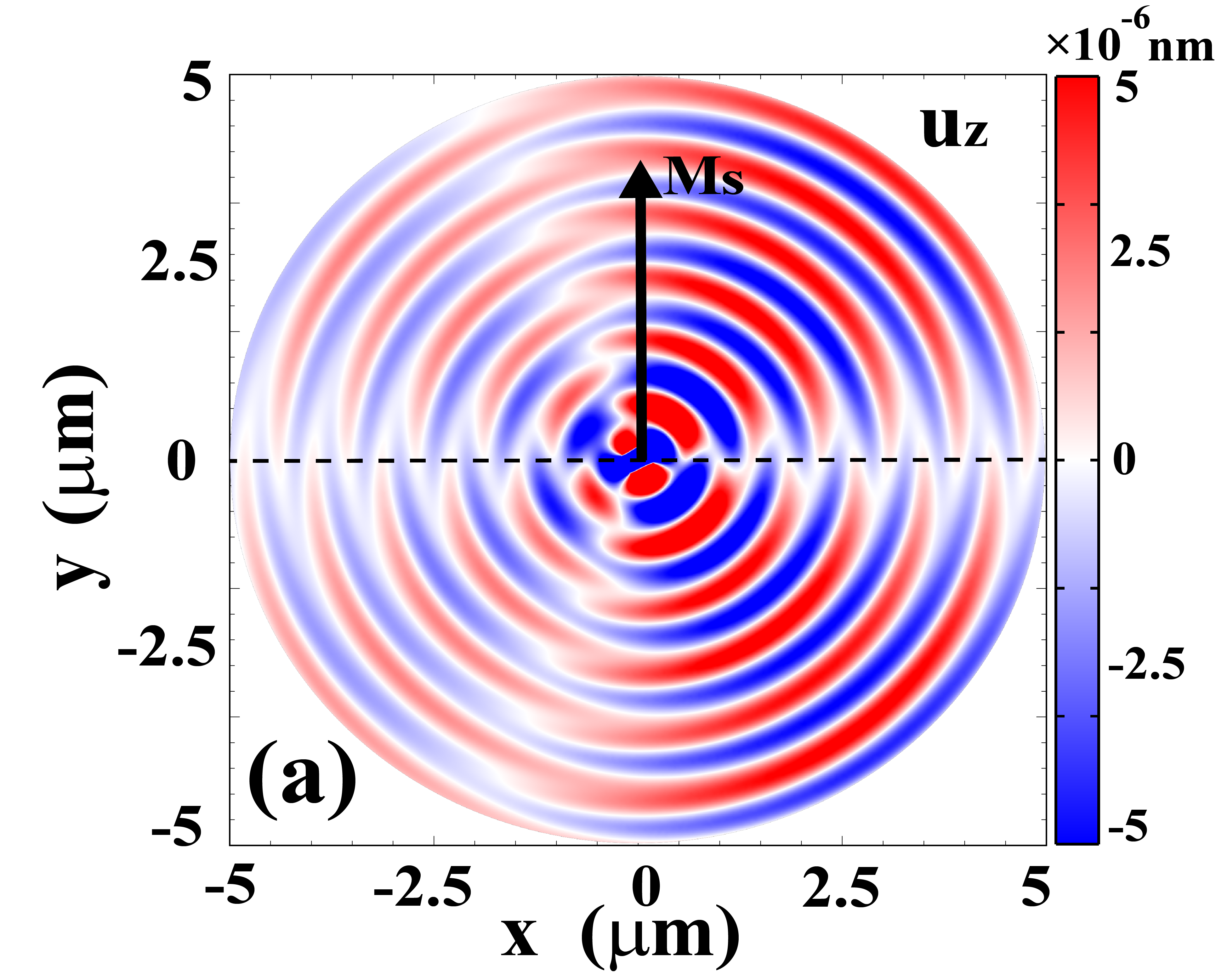}
\hspace{-0.1cm}\includegraphics[width=0.495\textwidth,trim=0.1cm 0.1cm 0.15cm 0.5cm]{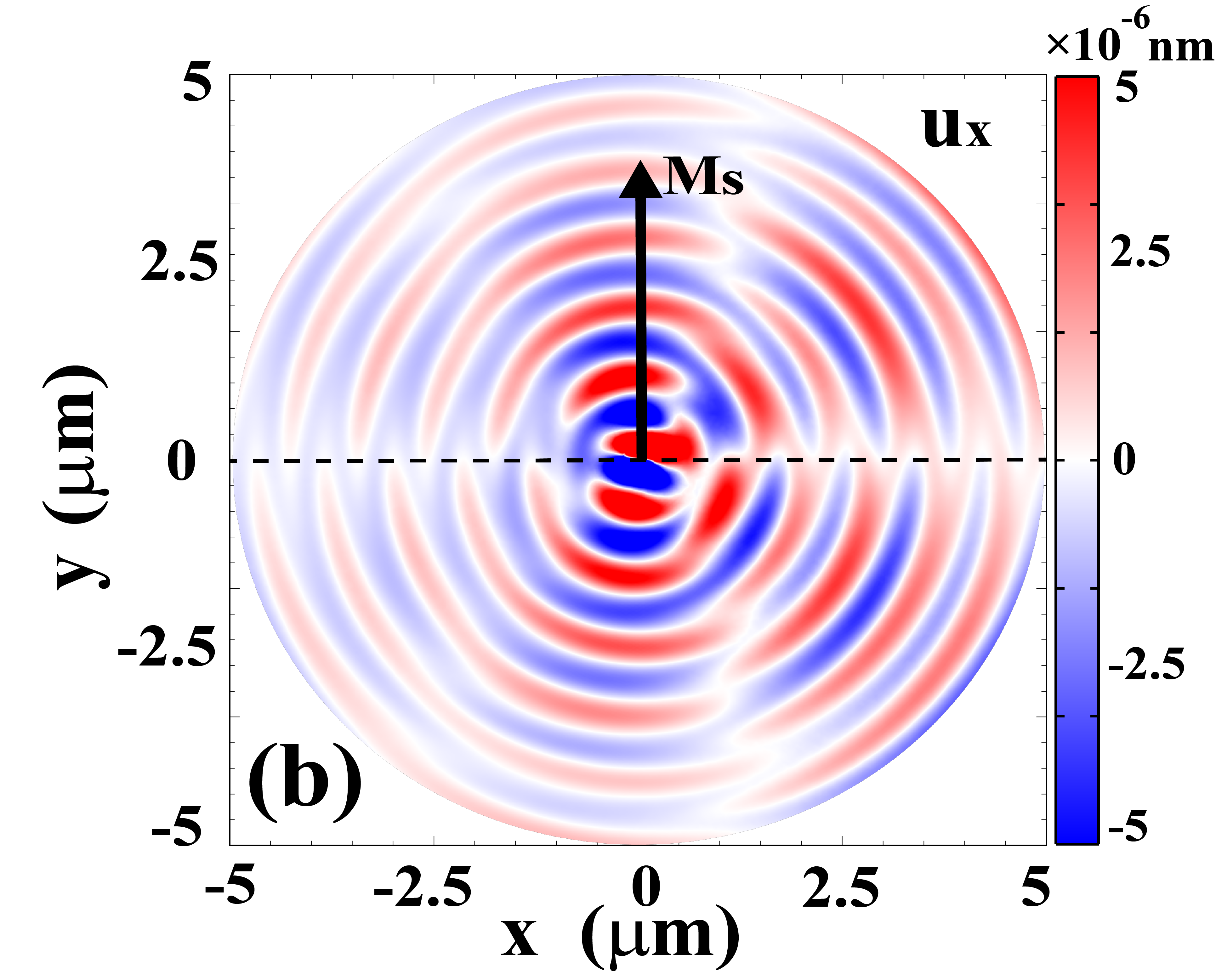}
\caption{Pumped displacement fields $u_z$ [(a)] and $u_x$ [(b)] at the surface $z=0$ of GGG substrate by the FMR of YIG disc of thickness $d=80$~nm and radius $r=150$~nm, which is saturated along the $\hat{\bf y}$-direction. The dashed line indicates the pattern with pure SHG.}
\label{magnetic_disc}
\end{figure}

 \textit{Directional SHG by magnetic nano-disc}.---It is convenient to resolve the above direction-dependent phonon pumping by a magnetic nano-disc. We consider a YIG disk of thickness $d=80$~nm and radius $r=150$~nm on the GGG substrate, with the magnetization biased along the $\hat{\bf y}$-direction by a static magnetic field $\mu_0H_0=0.1$~T. The demagnetization factor $N_{xx}=N_{yy}\simeq d/(2d+\sqrt{\pi}r)$, and $N_{zz}\simeq \sqrt{\pi}r/(2d+\sqrt{\pi}r)$ \cite{cylinder}. We apply a similar magnetic field pulse centered at frequency $\omega_{\rm F}=4$~GHz to the wire case such that the excited transverse magnetization $M_{z}=0.15M_s$. Figure~\ref{magnetic_disc} shows the pumped displacement fields $u_z$ and $u_x$ at the surface $z=0$ of the substrate. There exists a special direction denoted by the dashed line that exhibits pure SHG without any linear and third harmonics when the pumped SAWs propagate normally to ${\bf M}_s$ direction, similar to that by the magnetic wires.  The SHG mixes with the linear phonon pumping, however, when the SAWs propagate in the other directions.

  \textit{Conclusion.}---In conclusion, we predict the acoustic frequency comb with frequency multiplication of SAWs by magnetic transducers when driven by microwaves. We further predict the conditions to realize the pure acoustic SHG without any linear and third harmonics, a functionality beyond those by anharmonic interaction of lattice \cite{SHG_1,SHG_2,ultrasonic_1,nature_electronics}. Such a magnetic approach may overcome the difficulty in the electric technique in coherent phonon generation since it allows high-frequency ($>10$~GHz) excitation of phonons by microwaves with ultra low energy consumption and unprecedented tunability with different magnetic configurations and material choices, thus particularly useful in miniaturized phononic, magnonic, and spintronic devices.

  \vskip0.25cm	
\begin{acknowledgments}
This work is financially supported by the National Natural Science Foundation of China under Grant No.~0214012051, and the startup grant of Huazhong University of Science and Technology (Grants No.~3004012185 and No.~3004012198). W.Y. is supported by National Natural Science Foundation of China under Grant No.~12204107, and Shanghai Science and Technology Committee (Grants No.~21PJ1401500 and No.~21JC1406200).
\end{acknowledgments}

\end{document}